\documentclass[twocolumn,aps,prb,superscriptaddress,nofootinbib]{revtex4-1}
\usepackage{graphicx}
\usepackage{times}
\usepackage{bm}
\usepackage[linktocpage=true]{hyperref}
\usepackage{pgfplots}
\usepackage{amsmath}
\usepackage{braket}
\usetikzlibrary{arrows}
\usepackage{enumerate}
\usetikzlibrary[patterns] % ConTEXt when using Tik Z
\usetikzlibrary[snakes] % ConTEXt when using Tik Z
\usetikzlibrary[shapes] % ConTEXt when using Tik Z
\usepackage{pgfplots}
\usepackage{newfloat}
\DeclareFloatingEnvironment[name={Supplementary Figure}]{suppfigure}

\begin{document}

\title{Building topological quantum circuits: Majorana nanowire junctions}
\author{Tudor D. Stanescu}
\affiliation{Department of Physics and Astronomy, West Virginia University, Morgantown, WV 26506}
\affiliation{Condensed Matter Theory Center and Joint Quantum Institute, Department of Physics, University of Maryland, College Park, Maryland, 20742-4111, USA}
\author{Sankar Das Sarma}
\affiliation{Condensed Matter Theory Center and Joint Quantum Institute, Department of Physics, University of Maryland, College Park, Maryland, 20742-4111, USA}

\begin{abstract}
Topological quantum computation using non-Abelian Majorana zero modes localized in proximitized semiconductor nanowires requires careful electrostatic control of wire-junctions so as to manipulate and braid the zero modes enabling anyonic fault-tolerant gate operations.  We theoretically investigate the topological superconducting properties of such elementary wire-junctions, the so-called T-junctions, finding that the existence of the junction may nonperturbatively affect the Majorana behavior by introducing spurious non-topological subgap states mimicking zero-modes.  We propose a possible solution to this potentially serious problem by showing that junctions made lithographically from two-dimensional (2D) electron gas systems may manifest robust subgap topological properties without any spurious zero modes.  We propose a 2D structure that enables multiprobe tunneling experiments providing position-dependent spectroscopy, which can decisively settle outstanding open questions related to the origin of the zero-bias conductance peaks observed experimentally. We also find that junctions with trivial superconductors may result in local perturbations that induce extrinsic low-energy states similar to those associated with wire junctions. 
\end{abstract}

\maketitle

\section{Introduction}

Recent improvements in materials science, nanofabrication, and measurements \cite{Zhang2017,Nichele2017,Zhang2018} have led to the observation of stable zero energy subgap states in semiconductor nanowires proximity-coupled to superconductors. These states manifest the predicted $2e^2/h$ quantization of the zero bias differential conductance at low temperatures, as expected \cite{Sau2010a,Lutchyn2010,Sau2010,Oreg2010,Lutchyn2011,Stanescu2011} for topological non-Abelian Majorana zero modes (MZMs).  Questions, however, remain whether the observed zero bias conductance peaks (ZBCPs) arise from non-Abelian topological MZMs or from accidental non-topological subgap Andreev bound states (ABSs) \cite{Liu2017a,Moore2018,Kells2012,Lee2014,Klinovaja2015,Deng2016}, which are ubiquitous even in clean ballistic nanowires due to possible variations of the  electrochemical potential.  These accidental ABSs can be thought of as strongly overlapping MZMs  due to a smooth background potential, with one of the MZMs coupling strongly to the tunneling lead and producing a ZBCP that mimics the pure MZM behavior.  Thus, in spite of enormous advances in the Majorana nanowire experiments during 2012-2017, the central question of whether the theoretically predicted non-Abelian MZMs actually exist in proximitized nanowires remains open. Moreover, this question cannot be entirely settled through local tunneling  measurements, since the observation of a $2e^2/h$ zero bias tunneling conductance peak is only a necessary condition for the existence of MZMs.  Ultimately, a braiding experiment manifesting the non-Abelian nature of these MZMs is required.  Such MZM braiding may also be exploited for building a fault-tolerant topological quantum computer using MZMs for topologically-protected gate operations.

The next step in this exciting subject is, therefore, some kind of braiding experiment involving the MZMs to directly verify their non-Abelian topological properties.  Since the ABSs are by definition non-topological, they will have trivial braiding properties, thus providing a sufficient condition to distinguish between MZM and ABS.  Many different architectures for MZM braiding have been proposed \cite{Alicea2011,Sau2011,Aasen2016,Karzig2017,Litinski2017}.  These proposals, however, involve wire junctions where two nanowires or a nanowire and a trivial superconductor come together.  There are various such junctions (e.g., T-junctions, Y-junctions \cite{Lutchyn2017}, hash-tags \cite{Gazibegovic2017}) that have been proposed in this context, with the basic idea being that electrostatic operations using external  gates would enable an effective ``movement'' of the MZMs by creating and eliminating them from various wire ends.  The details of how these operations are carried out are of no significance in the current work. Here, we ask a very simple (and truly basic) question:  Given a junction, where two nanowires come together, how does the existence of the ``passive wire'' affect the low lying MZMs in the ``Majorana wire''?  Rather surprisingly, this very elementary (and obviously crucial) question has not been addressed in the literature, in spite of wire-junction braiding structures being the cornerstone of all proposed Majorana nanowire-based topological circuits \cite{Alicea2011,Karzig2017}.

In the current paper, we investigate theoretically the low-lying subgap electronic structure of the Majorana wire in the presence of a passive wire within the minimal model used extensively for modeling Majorana nanowire systems. We find that the wire-junction system exhibits a low-lying subgap structure that may be qualitatively different from that of a simple Majorana nanowire. This happens  due to the junction itself, without assuming the existence of any other problematic ABS mode that may exist in the Majorana wire itself (e.g., due to nonuniform electrochemical potentials \cite{Liu2017a,Moore2018}).  We show that there are unexpected nonperturbative effects of the passive wire on the Majorana wire, which may completely compromise the topological protection of the Majorana subspace and lead to serious complications in  the low-energy spectral properties of the Majorana wire when using the passive wire as a tunneling probe.
  In particular, spurious zero modes may arise in the Majorana wire, unless special care is taken in the design of the wire-junction.  We provide practical guidelines for the construction of junctions that preserve the intrinsic low-energy properties of the Majorana wire and critically discuss some pitfalls to avoid in building topological quantum computing (TQC) circuits. 

A key goal of this work is to elucidate the current experimental situation underlying MZM observations, where both trivial ABSs and topological MZMs lead to similar zero bias peaks (ZBPs) in end-of-wire tunneling conductance spectroscopy \cite{Liu2017a}.  
Obviously, any serious effort to building TQC circuits must first clarify the ABS-MZM dichotomy and ensure that the only low-lying states in the system are the spatially isolated Majorana zero-energy modes localized at the ends of the wire. 
The simplest test that would corroborate the MZM interpretation of the observed ZBPs is the observation of the closing (and re-opening) of the bulk gap at the topological quantum phase transition (TQPT). Note that a reopening of the gap with increasing magnetic field has never been clearly observed in Majorana nanowire experiments.  This observation  requires tunneling into the bulk of the wire \cite{Stanescu2012}, hence a T-wire geometry where the passive wire is used as a probe.  Another simple (but crucial) necessity is the observation of correlated ZBPs at the two ends of the wire, which arise from the nonlocal topological nature of the MZMs \cite{DSarma2012}. 
A T-wire geometry would enable both these experiments, which are crucial for establishing whether or not the nanowires are actually suitable for enabling more complicated interferometric, fusion, or braiding experiments. %thus paving the way for the demonstration of MZMs in these systems. 

Our careful consideration of the nonperturbative physics of T-junctions leads to two important results: (1) We show that naive implementations of T-wire  structures result in strong perturbations that  induce additional low-energy states. These states  may corrupt the results of a tunneling measurement and destroy the topological properties of the system.  (2)  Based on our theoretical analysis, we propose a 2D structure that solves  the ``strong perturbation'' problem for the observations of both TQPT and correlated ZBPs.  In fact, our proposed  2D structure enables a multiprobe tunneling experiment  that provides position-dependent spectroscopy, which can decisively settle the issues related to ZBP-correlations and bulk-gap-reopening, as well as detect possible low-energy ABSs localized inside the wire. This multiprobe technique could be used as a testing tool  in the fabrication of future TQC circuits.

The rest of the paper is organized as follows. In Sec. II we investigate the low-energy properties of T-junctions between two proximitized semiconductor wires and of junctions made lithographically from 2D electron gas systems hosted by semiconductor heterostructures.  We also propose a specific 2D structure that enables multiprobe tunneling experiments capable of providing position-dependent spectroscopy. In Sec. III we discuss  
junctions between proximitized semiconductor wires and trivial superconductors. We also analyze
a simple solution for grounding a topological superconducting island using a standard T-junction. Our concluding remarks are presented in Sec. IV.

\section{Wire junctions and 2D structures}

\begin{figure}[t]
\begin{center}
\includegraphics[width=0.46\textwidth]{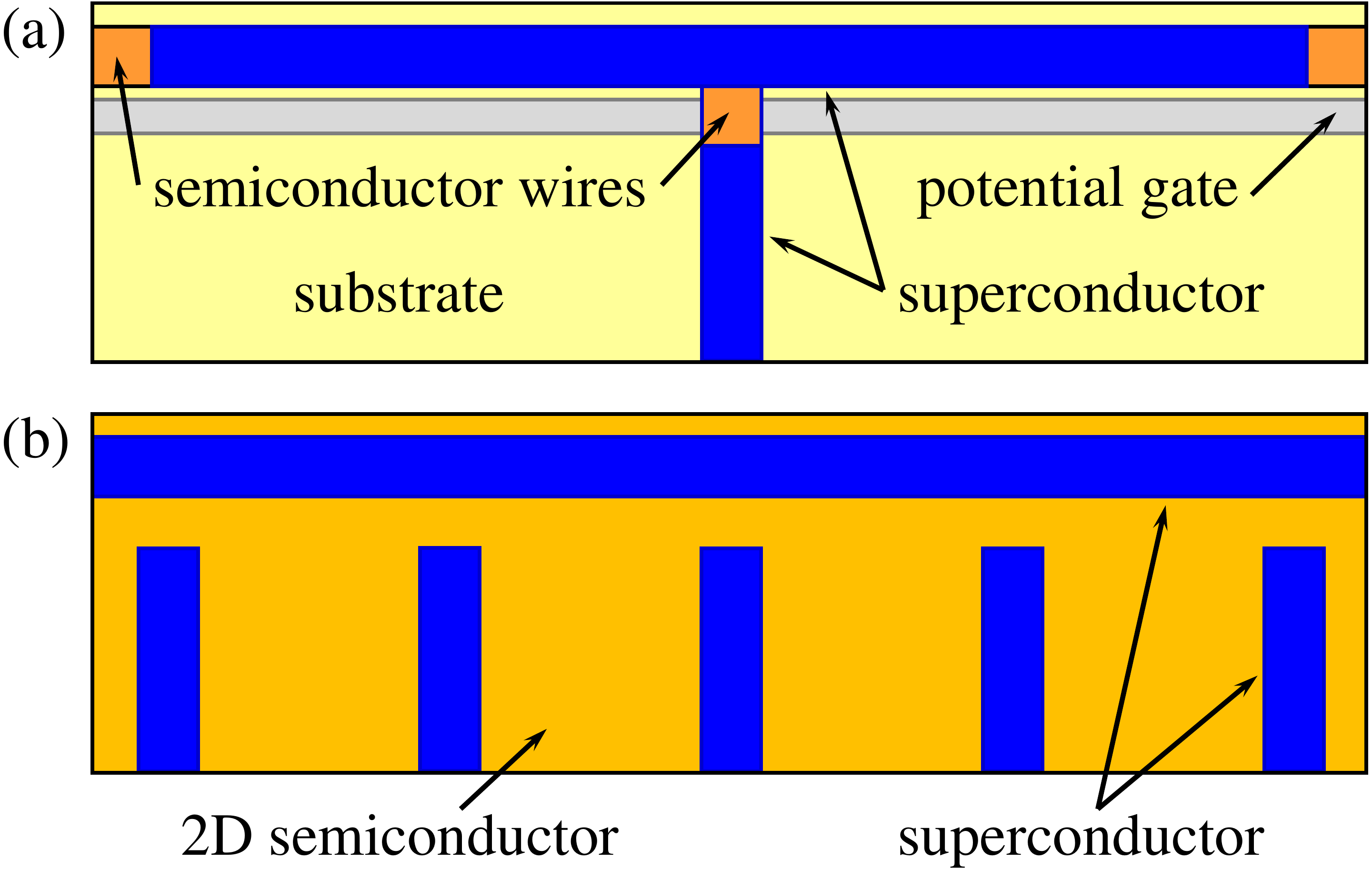}
\end{center}
\vspace{-3mm}
\caption{(Color online) Schematic representation of the SM-SC devices studied in this work (top view). (a) Proximitized semiconductor wires in the T-wire geometry. The vertical wire can be used to probe the LDOS in the middle of the Majorana wire by measuring the differential conductance for tunneling through a potential barrier generated by a narrow gate. (b) Two-dimensional SM-SC device in the ``piano keyboard'' geometry. A 2D electron gas hosted by a semiconductor heterostructure is partially depleted using a top gate (not shown), except along quasi-1D wires defined by the superconductor pattern deposited on the semiconductor.}
\label{FIG1}
\vspace{-3mm}
\end{figure}

A T-wire semiconductor-superconductor (SM-SC) device, which could be used to probe the local density of states (LDOS) in the middle of a Majorana wire and could demonstrate the closing (and re-opening) of the bulk gap at the TQPT, is represented schematically in Fig. \ref{FIG1}(a). 
The horizontal segment (the ``Majorana wire'') represents the hybrid system that hosts the MZMs, while the vertical (``passive'') wire is used as a tunneling probe.  A  narrow back gate creates a tunnel barrier at the junction, while a magnetic field is applied parallel to the horizontal (Majorana) wire. A two-dimensional (2D) device allowing the measurement of the LDOS at several different locations along the Majorana wire is represented in   Fig. \ref{FIG1}(b). The semiconductor heterostructure hosts a 2D electron gas that is partially depleted by applying a top gate, except along the pattern defined by the superconductor. A magnetic field applied parallel to the horizontal wire drives it into a topological SC phase, while its LDOS at several different locations  is probed by tunneling through the vertical wires. 

%%%%%%%%%%%%%%%%%%%%%%%%%%%%%%%%
%%%%%%%%%%%%%%%%%%%%%%%%%%%%
%%%%%%%%%%%%%%%%%%%%%%	
%%%%%%%%%%%%%%%%%%%%%%%%%%%%%%%%

\begin{figure}[t]
\begin{center}
\includegraphics[width=0.46\textwidth]{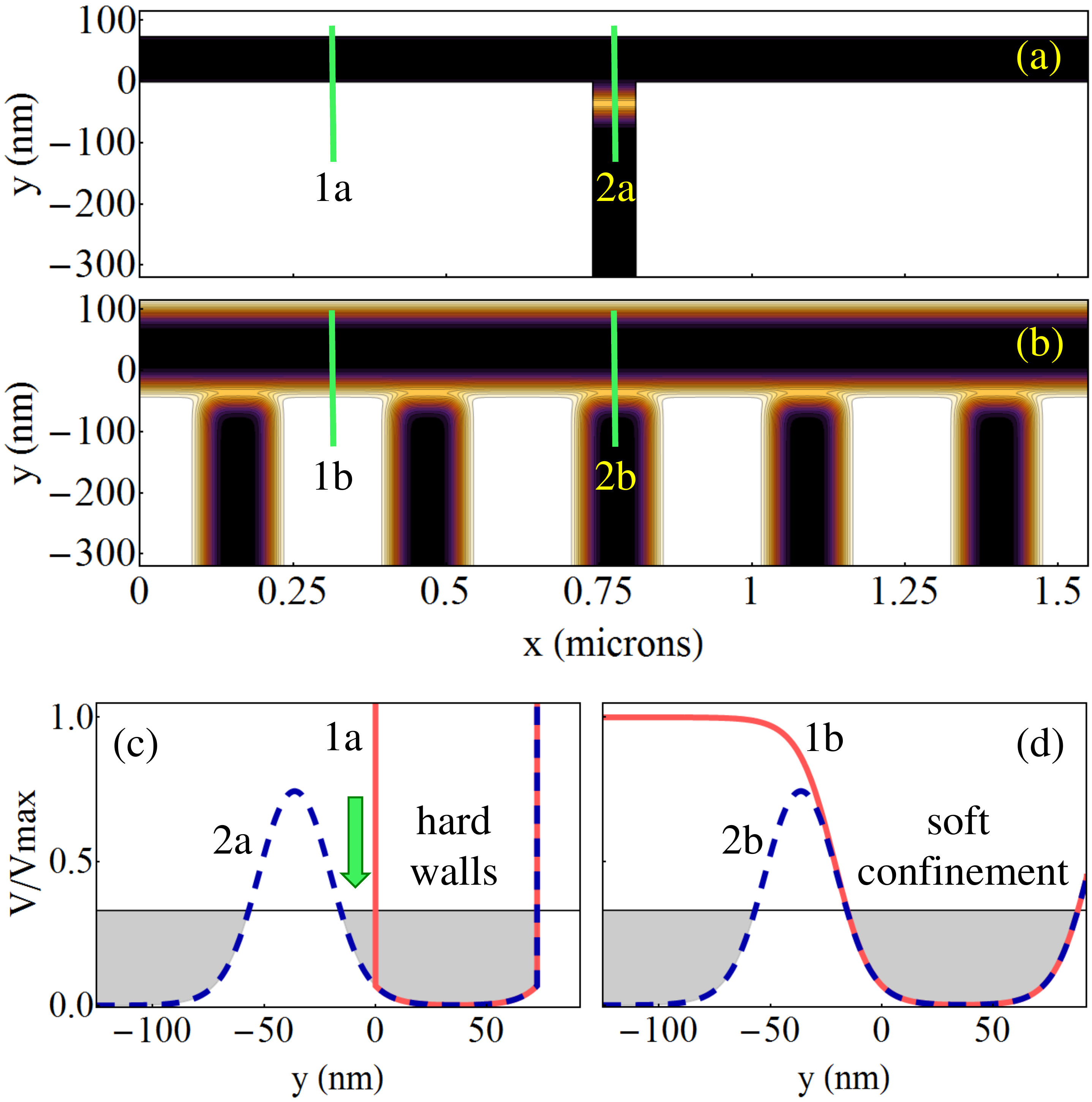}
\end{center}
\vspace{-3mm}
\caption{(Color online)  Effective confining potential for the devices shown in Fig. \ref{FIG1}. (a) Confining potential for SM wires in the T-wire geometry.  The potential profiles along the cuts `1a' and `2a' are shown in panel (c). (b) Effective potential for the 2D structure from Fig. \ref{FIG1}(b). 
Black corresponds to zero potential, while white represents regions with a potential equal to [or, in panel (a), higher than] a certain value $V_{\rm max}$. The potential profiles along the cuts `1b' and `2b' are shown in panel (d). The thin horizontal lines in panels (c) and (d) represent the chemical potential.}
\label{FIG2}
\vspace{-3mm}
\end{figure}

The impact of a junction on the low-energy physics of the Majorana wire can be physically understood by analyzing the profile of the effective confining potential in the system.  Fig. \ref{FIG2} shows the potential profiles for the devices represented schematically in Fig. \ref{FIG1}.  For the T-wire device [panels (a) and (c)], the system is characterized by hard-wall transverse confinement [red line in panel (c)] everywhere except the junction region  [dashed blue line in panel (c)], where the potential barrier provides a ``soft'' confinement. This soft confinement allows electrons from the Majorana wire to partially penetrate into the passive wire, as suggested by the green arrow in panel (c). The weaker lateral confinement in the junction region may represent a significant perturbation for the system, as we explicitly show below. 
By contrast, in the 2D structure [panels (b) and (d)] the transverse confinement is everywhere ``soft''  and essentially position-independent [see panel (d)].  The uniformity of the effective potential in the Majorana wire means that the probes do not perturb the system significantly. Consequently, the differential conductance  measured by the probes represents an intrinsic property of the Majorana wire (roughly proportional to the LDOS near the corresponding junction). On the other hand, the low-energy properties measured using the T-wire device include additional features generated by the strong perturbation induced by the junction itself and thus are not intrinsic properties of the Majorana wire.

%%%%%%%%%%%%%%%%%%%%%%%%%%%%%%%%
%%%%%%%%%%%%%%%%%%%%%%%%%%%%
%%%%%%%%%%%%%%%%%%%%%%	
%%%%%%%%%%%%%%%%%%%%%%%%%%%%%%%%

We investigate the low-lying electronic structure of the Majorana wire in the presence of one (or more) passive wire(s) using the minimal tight-binding model extensively used for modeling Majorana nanowire systems \cite{Stanescu2013}. The Bogoliubov - de Gennes (BdG) Hamiltonian, which includes the ``standard terms'', i.e., nearest neighbor hopping, Zeeman splitting, Rashba spin-orbit coupling, and induced pairing,  is defined on a square lattice with lattice constant $a$.  The effective Hamiltonian describing the low-energy electronic properties of the semiconductor-superconductor (SM-SC) structure has the form:
\begin{eqnarray}
H_{\rm eff} &=& -t\sum_{{\bm i},{\bm \delta}}c_{\bm i}^\dagger c_{{\bm i}+{\bm \delta}}+\sum_{\bm i}[\mu+\widetilde{V}({\bm i})]c_{\bm i}^\dagger c_{\bm i} + \Gamma\sum_{\bm i}c_{\bm i}^\dagger\hat{\sigma}_x c_{\bm i} \nonumber \\
 &+& \frac{i\alpha_R}{2}\sum_{\bm i}\left(c_{{\bm i}+{\bm \delta}_x}^\dagger \hat{\sigma}_y c_{{\bm i}} - c_{{\bm i}+{\bm \delta}_y}^\dagger \hat{\sigma}_x c_{{\bm i}} + {\rm h.c.}\right)       \label{Heff} \\
 &+& \Delta\sum_{\bm i} \left(c_{{\bm i}\uparrow}^\dagger c_{{\bm i}\downarrow}^\dagger +  c_{{\bm i}\downarrow}c_{{\bm i}\uparrow}\right), \nonumber
\end{eqnarray}
where $c_{\bm i}^\dagger =(c_{{\bm i}\uparrow}^\dagger, c_{{\bm i}\downarrow}^\dagger)$ is the electron creation operator on the
site ${\bm i}= (i_x, i_y)$ of a square lattice  with lattice constant $a$, ${\bm \delta}_x=(1,0)$, ${\bm \delta}_y=(0,1)$, and ${\bm \delta} = \pm{\bm \delta}_x$ or $\pm {\bm \delta}_y$ are nearest neighbor vectors, and $\hat{\sigma}_\mu$, with $\mu= x,~y,~z$ are Pauli matrices associated
with the spin degree of freedom. The numerical calculations shown below are done on a lattice with $a=10~$nm, but we have checked that the results are basically the same for $a=5~$nm. 
The model parameters characterizing the effective Hamiltonian are the nearest-neighbor hopping, $t=1270/a^2~$meV (where $a$ is given in nanometers) corresponding to an effective mass $m_{\rm eff} = 0.03 m_0$, where $m_0$ is the bare electron mass, the Zeeman field, $\Gamma$ (taking values in the range $0$ to $1.25~$meV), the Rashba spin-orbit coupling, $\alpha_R=25/a~$meV (which corresponds to $250~$meV$\cdot$\AA), and the  induced pairing, $\Delta=0.25~$meV. The chemical potential $\mu$ was set near the bottom of the second confinement-induced band of the Majorana wire, i.e. $\mu\approx 5.8~$meV for the T-wire structure and $\mu\approx 5~$meV for the 2D system, while the confining potential $\widetilde{V}({\bm i}) = V(a i_x, a i_y)$ corresponds to the profiles shown in Fig. \ref{FIG2}. The Majorana wire from Fig. \ref{FIG1}(a) is modeled by seven parallel chains of length $N_x=155$, which corresponds to a wire of width $L_y=70~$nm and length $L_x=1.55~\mu$m. A $200~$nm long segment of the passive wire is also included in the calculation, with a $70~$nm junction region that is not covered by the superconductor.  The 2D device from Fig. \ref{FIG1}(b) is modeled on a lattice consisting of 25 parallel chains of length $N_x=155$. The quasi-1D wires are defined by the effective potential shown in Fig. \ref{FIG2}(b). This confining potential is obtained under the assumption that the superconductor pattern screens the applied top gate potential with a characteristic screening length of $40~$nm. 
We note that studying 2D structures can be numerically challenging even within simple models, as it involves a large number of degrees of freedom. In our case, the effective Bogoliubov -- de Gennes problem has dimension $4 N_x N_y$. To optimize the numerical analysis, we project the problem onto a low-energy subspace defined by the eigenstates of the first two terms in Eq. (\ref{Heff}) (i.e. semiconductor with no spin-orbit coupling, Zeeman splitting, and induced pairing, but in the presence of the confining potential)  with energies lower than $\Delta E = 35~$meV. 
  Note that we do not include any intrinsic ABSs in the Majorana wire.
 
\begin{figure}[t]
\begin{center}
\includegraphics[width=0.44\textwidth]{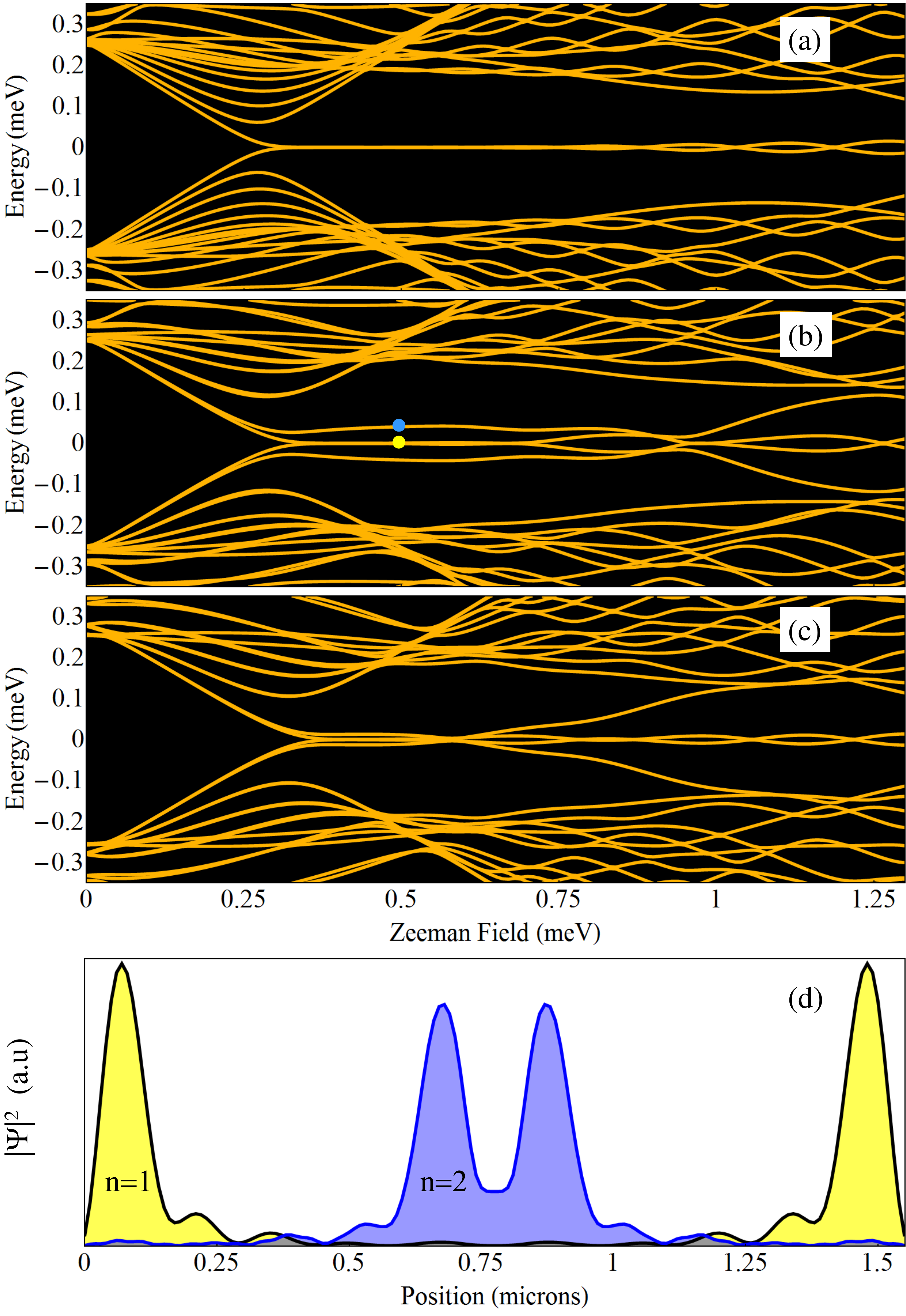}
\end{center}
\vspace{-4mm}
\caption{(Color online) (a) Low-energy BdG spectrum as a function of the Zeeman field for a proximitized wire that is not coupled to a probe -- i.e., no vertical segment in Fig. \ref{FIG1}(a). (b) Low-energy spectrum for the T-wire geometry with an effective confinement as shown in Figs. \ref{FIG2}(a) and \ref{FIG2}(c) and $V_{\rm max}=15~$meV.  (c) Same as panel (b) for $V_{\rm max}=20~$meV. (d) Wave functions of the lowest energy states marked by small circles in panel (b). The bound state $n=2$ is generated by the perturbation introduced by the junction.}
\label{FIG3}
\vspace{-3mm}
\end{figure}

The dependence of the low-energy spectrum of the 1D Majorana wire on the applied Zeeman field is shown in Fig. \ref{FIG3}. Panel (a) shows the unperturbed spectrum that characterizes the wire in the absence of a probe. Note that the bulk bands have a minimum at a ``critical'' field $\Gamma_c\approx 0.3~$meV corresponding to the finite-size remnant of the TQPT. For $\Gamma>\Gamma_c$, a Majorana mode emerges in the middle of the topological gap.  The presence of a probe perturbs the Majorana wire, which may result in additional low-energy states. This is illustrated in Fig. \ref{FIG3} panels (b) and (c) for a T-wire structure with an effective potential as shown in Fig. \ref{FIG2}(a) and (c). The height of the potential barrier between the wire and the probe is $V_b\approx 11.25~$meV in panel (b) and $V_b\approx 15~$meV in panel (c). 
The additional low-energy mode that emerges above the critical field does not represent an intrinsic property of the Majorana wire but is rather the result of coupling to the probe.
We note that, in general, the strength of the perturbation induced by the junction depends on the details of the structure, including the height and width of the potential barrier, chemical potential,  effective diameter of the probe in the junction region (which may be different from the diameter of the main wire), etc. The extrinsic low-energy states induced by this perturbation are Andreev bound states (ABSs)  localized in the middle of the Majorana wire. An example of such a state is shown in panel (d). We have verified that the basic physics of junction-induced ABSs  is a generic property of wire  junctions for reasonable representative model parameters. More realistic modeling (i.e. beyond the generic model used in this study) may be necessary to understand the quantitative details of sub-gap features induced by wire junctions and to optimize devices with specific geometries and materials composition.
 
 %%%%%%%%%%%%%%%%%%%%%%%%%%%%%%%%
%%%%%%%%%%%%%%%%%%%%%%%%%%%%
%%%%%%%%%%%%%%%%%%%%%%	
%%%%%%%%%%%%%%%%%%%%%%%%%%%%%%%%
 
The emergence of perturbation-induced low-energy states in wire junctions raises a serious problem regarding the possible use of these structures in topological quantum circuits. What makes the situation even worse is that the specific properties of these states (e.g., the dependence on the Zeeman field) are extremely sensitive to the the details of the junction, making it virtually impossible to predict their exact behavior (or to figure out whether particular low-energy states in the system are trivial or topological). 
This feature  also suggests that using the T-wire geometry to probe the LDOS in the middle of the wire may not be a useful approach. One could never disentangle the intrinsic properties of the Majorana wire from the perturbation-induced features. In this context, note that the bulk bands associated with the closing of the gap at the TQPT are also significantly perturbed, as evident when comparing panels (b) and (c) with panel (a) in Fig. \ref{FIG3}.

%%%%%%%%%%%%%%%%%%%%%%%%%%%%%%%%
%%%%%%%%%%%%%%%%%%%%%%%%%%%%
\begin{figure}[t]
\begin{center}
\includegraphics[width=0.45\textwidth]{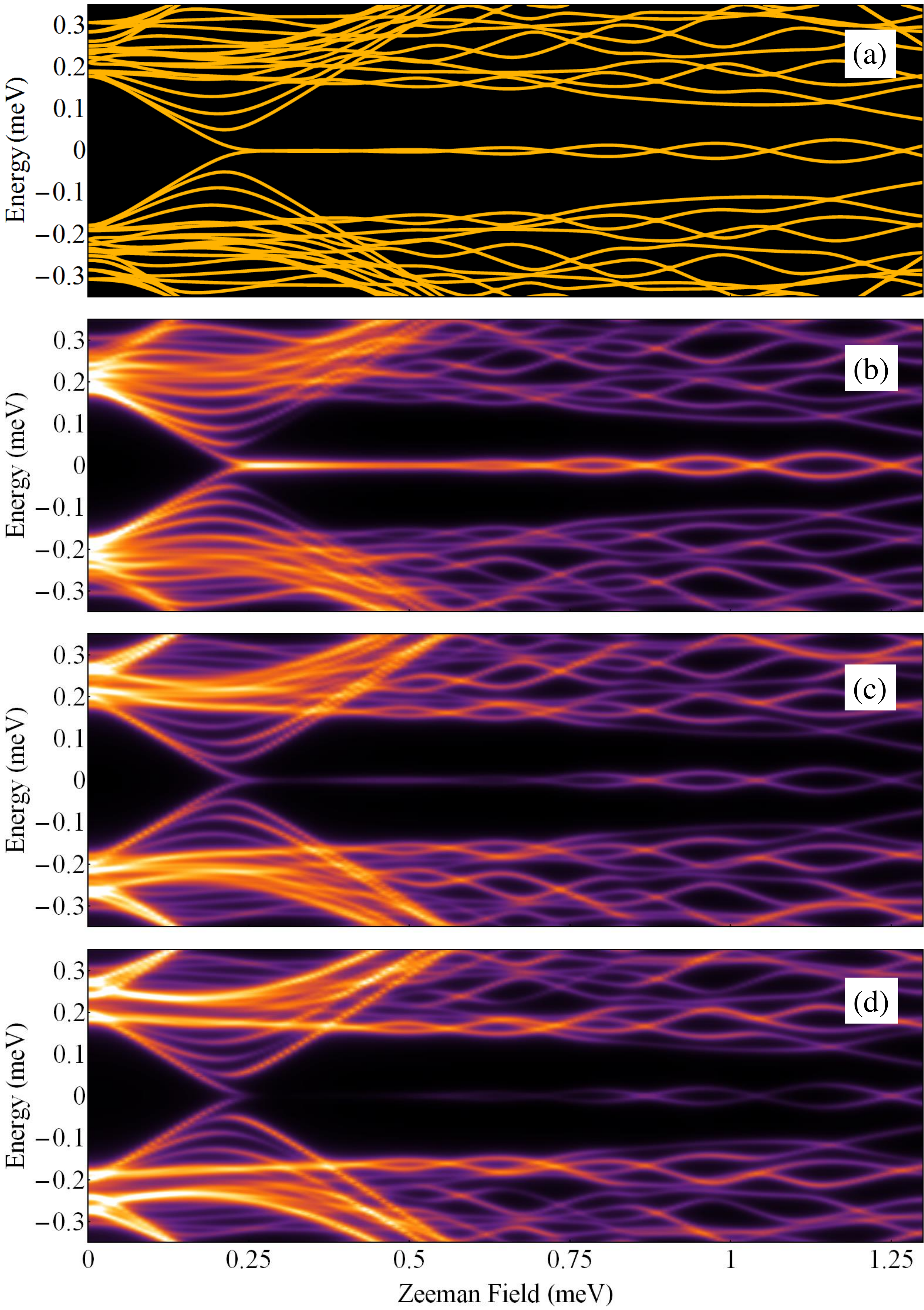}
\end{center}
\vspace{-5mm}
\caption{(Color online) Low-energy spectrum and LDOS as functions of the Zeeman field for the 2D structure shown in Fig. \ref{FIG1}(b). (a) Low-energy BdG spectrum for the unperturbed wire.  (b)--(d) LDOS at the ends of the first three probes from Fig. \ref{FIG1}(b). In the low-tunneling regime, these local densities of states are expected to be proportional to the corresponding differential conductances. The effective confining potential of the 2D structure is shown in Figs. \ref{FIG2} (b) and \ref{FIG2}(d) and $V_{\rm max}=15~$meV.}
\label{FIG4}
\vspace{-3mm}
\end{figure}
%%%%%%%%%%%%%%%%%%%%%%	
%%%%%%%%%%%%%%%%%%%%%%%%%%%%%%%%

To address these problems, we suggest using the 2D semiconductor-superconductor structure \cite{Suominen2017,Hell2017,Hell2017a,Pientka2017} shown in Fig. \ref{FIG1}(b). The key element in this device is that the effective confinement of the Majorana wire is weakly position-dependent, as discussed in the context of Fig. \ref{FIG2}. This means that the perturbation induced by the probes is negligible and the measured differential conductance actually reflects the intrinsic properties of the Majorana wire. The results of a calculation supporting these claims are shown in Fig. \ref{FIG4}. Panel (a) represents the low-energy spectrum of an unperturbed wire.  This spectrum 
 characterizes the intrinsic low-energy properties that, ideally, should be probed without perturbing the system. The LDOS at the ends of the first three probes, which is directly  measured in tunneling spectroscopy, is shown in panels (b)--(d).  The left-most probe [see panel (b)] shows a strong (nearly) zero-energy feature emerging above $\Gamma_c\approx 0.25~$meV. This feature is induced by the Majorana bound state localized near the left end of the system. The strength of this feature decreases in panels (c) (LDOS at the end of the second probe) and (d) (LDOS at the end of the middle probe) reflecting the (exponential) decay of the Majorana wave function away from the end of the wire. By contrast, the ``closing'' of the bulk gap associated with the  remnant TQPT is visible at the end of each probe, as the corresponding wave functions are delocalized. Note that placing the left-most probe at the very end of the system may reduce the visibility of the bulk states (which vanish at the boundaries of the system). Also note that the visibility of the bulk bands decreases with the size of the system, as the corresponding states are delocalized, hence their amplitude decreases with $L_x$. Finally, note that the strongest feature associated with the re-opening of the gap can be seen in panel (d), i.e. in the middle of the wire. All the features revealed by the LDOS can be traced back to the unperturbed spectrum shown in panel (a), which demonstrates that the perturbation introduced by the probes is, indeed, negligible.

%%%%%%%%%%%%%%%%%%%%%%%%%%%%%%%%
%%%%%%%%%%%%%%%%%%%%%%%%%%%%
\begin{figure}[t]
\begin{center}
\includegraphics[width=0.48\textwidth]{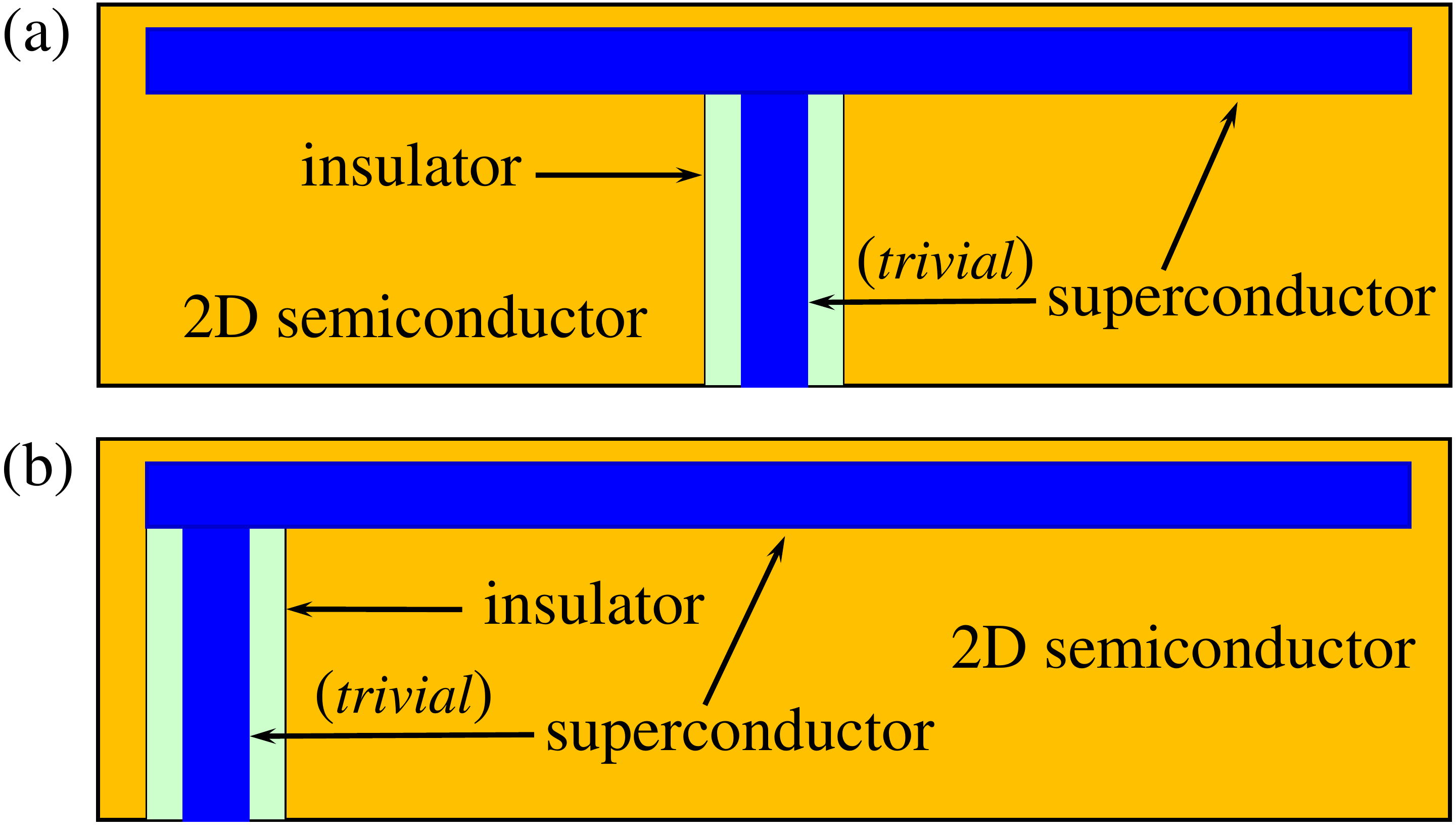}
\end{center}
\vspace{-2mm}
\caption{(Color online) Schematic representation of 2D structures consisting of a Majorana wire in contact with a trivial superconductor.  The trivial SC is deposited in a region of the 2D structure (the ``insulator'') that does not host a 2D electron gas. The electron gas in the orange region (i.e., uncovered semiconductor heterostructure) is depleted using a top gate (not shown).}
\label{FIG1S}
\vspace{-3mm}
\end{figure}
%%%%%%%%%%%%%%%%%%%%%%	
%%%%%%%%%%%%%%%%%%%%%%%%%%%%%%%%

\section{Junction with a trivial superconductor}

In this section we focus on junctions between wires and trivial superconductors and show that such a junction may result in a local
perturbation that induces extrinsic low-energy states similar to those associated with standard wire junctions. We also analyze
a simple solution for grounding a topological SC island using a standard T-junction and show that, although this induces trivial ABSs, in certain conditions the signature of these additional low-energy states is not visible in the tunneling conductance at the two ends of the wire, which can exhibit nonlocal
correlations. We note that three terminal Josephson T-junctions composed of  wires connected by a normal metal region have been studied in Ref. \onlinecite{Spanslatt2017}.

In an experiment involving the measurement of tunneling differential conductance at both ends of the wire the parent superconductor has to be grounded. If the SM-SC hybrid system is fabricated by sputtering the parent superconductor (e.g., NbTiN) on the semiconductor wire, this is not a problem since the superconductor is large, can be easily contacted, and covers the wire uniformly  (i.e. independent of the position {\em along} the wire). If, on the other hand, the superconductor is grown epitaxially (on a SM wire or a 2D semiconductor), it forms a relatively small island and  grounding it involves a junction with a trivial superconductor. This junction may induce a local perturbation strong-enough to generate unwanted low-energy states in the Majorana wire. Furthermore, this type of junction may be necessary to build topological qubits, e.g.,  when combining two or more wires into tetrons or hexons.\cite{Karzig2017,Plugge2017}  The key question is how to design a junction with a trivial superconductor in such a way as to avoid the creation of spurious low-energy states. 

Focusing on 2D structures, we consider a Majorana wire defined by an isolated SC stripe deposited on the semiconductor. The 2D electron gas hosted by the semiconductor heterostructure is depleted using a top gate everywhere except under the SC stripe. Grounding the Majorana island by simply connecting it with another SC stripe deposited on the semiconductor will prevent  the depletion of electrons under the contact, which  results in a T-type wire junction. One would  need to first ``remove'' the 2D electron gas in a certain region, then grow the (trivial) SC contact. This may involve serious technical challenges, but even if a solution is  found, one has to worry about the possibility of locally perturbing the Majorana wire and thus generating unwanted low-energy states. 

%%%%%%%%%%%%%%%%%%%%%%%%%%%%%%%%
%%%%%%%%%%%%%%%%%%%%%%%%%%%%
\begin{figure}[t]
\begin{center}
\includegraphics[width=0.48\textwidth]{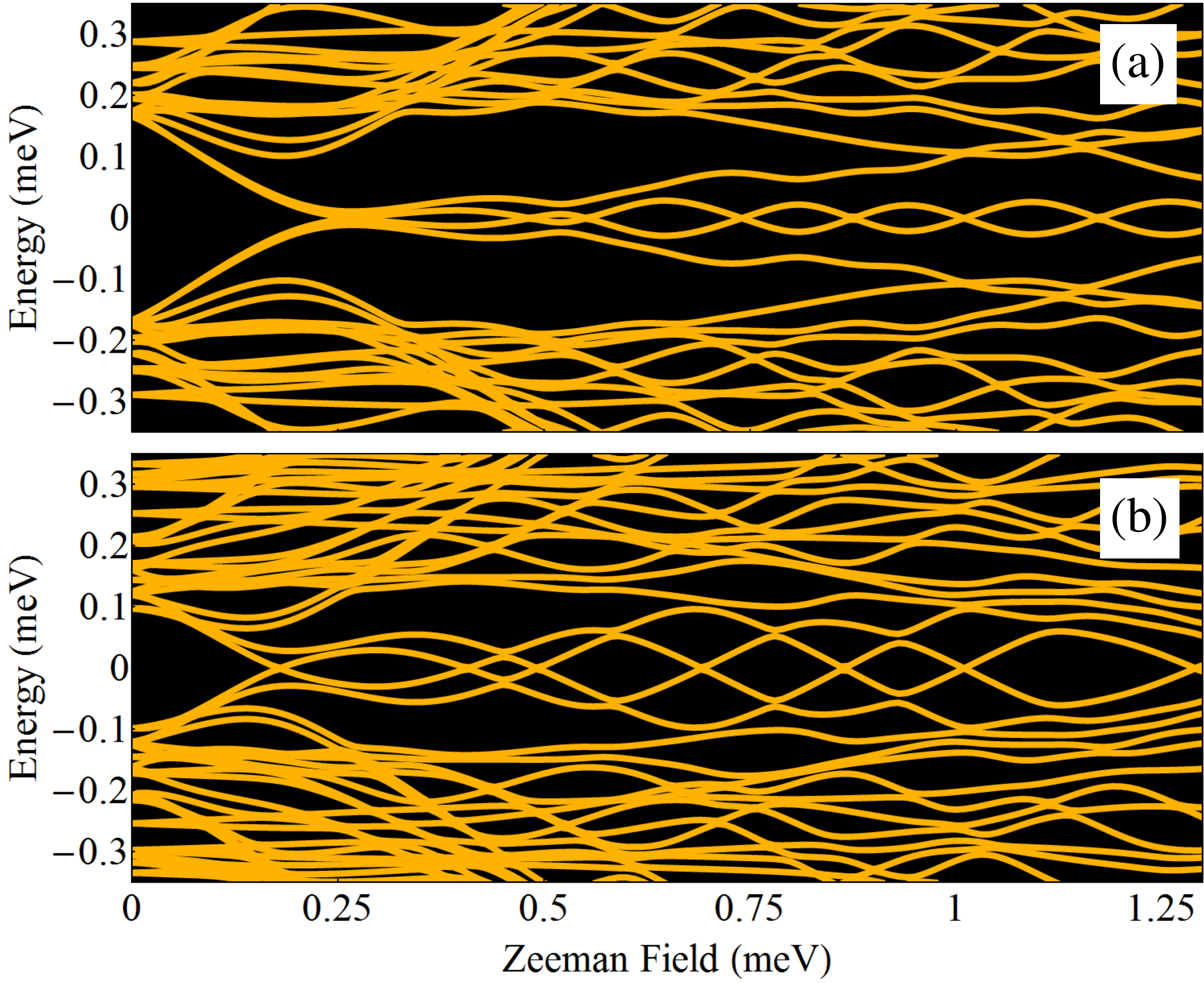}
\end{center}
\vspace{-2mm}
\caption{(Color online) Low-energy spectrum as a function of the applied Zeeman field for the 2D structure represented schematically in panel (a) of  Fig. \ref{FIG1S} (i.e. junction near the middle of the Majorana wire) and two different values of the chemical potential: (a) chemical potential near the bottom of the second confinement induced band ($\mu=3.6~$meV) and (b) chemical potential near the bottom of the third band ($\mu=8.1~$meV).}
\label{FIG2S}
\vspace{-3mm}
\end{figure}
%%%%%%%%%%%%%%%%%%%%%%	
%%%%%%%%%%%%%%%%%%%%%%%%%%%%%%%%

As an example, we consider the 2D structures represented schematically in Fig. \ref{FIG1S}. The trivial SC is deposited in a region of the 2D structure that does not host a 2D electron gas, which we dub the ``insulator''. In the numerical calculations we model the insulator as hard wall potential barrier.  
If the junction with the trivial SC is located near the middle of the Majorana wire, as shown in Fig. \ref{FIG1S} (a), it can induce unwanted low-energy states that may alter the results of tunneling measurements at the ends of the wire and compromise the topological protection of the Majorana subspace. This situation is illustrated in Fig. \ref{FIG2S}, which shows the dependence of the low-energy spectrum on the applied Zeeman field for two values of the chemical potential.  One can clearly see four low-energy modes, instead of the two intrinsic Majorana modes that characterize the Majorana wire. Note that the induced gap in panel (b) is significantly lower than the induced gap in panel (a). This is due to the fact that, for a given confining potential, the effective lateral confinement of the Majorana wire depends on the chemical potential. In particular, the states at the bottom of the second confinement-induced band [which are responsible for the low-energy features in panel (a)] are more confined under the SC stripe as compared to the  states at the bottom of the third band [panel (b)], which leak more into the uncovered SM region. The induced gap is larger for states that are more confined under the parent SC. We also note that, in addition to the dependence on the chemical potential illustrated in Fig. \ref{FIG2S}, the specific properties of the junction-induced low-energy states are strongly dependent on the details of the junction, e.g., width of the insulating region, distance between the insulator and the Majorana wire, etc. Nonetheless, our numerical analysis suggests that a device similar to that shown in Fig. \ref{FIG1S} is not an optimal solution for engineering a junction with a trivial superconductor.

%%%%%%%%%%%%%%%%%%%%%%%%%%%%%%%%
%%%%%%%%%%%%%%%%%%%%%%%%%%%%
\begin{figure}[t]
\begin{center}
\includegraphics[width=0.48\textwidth]{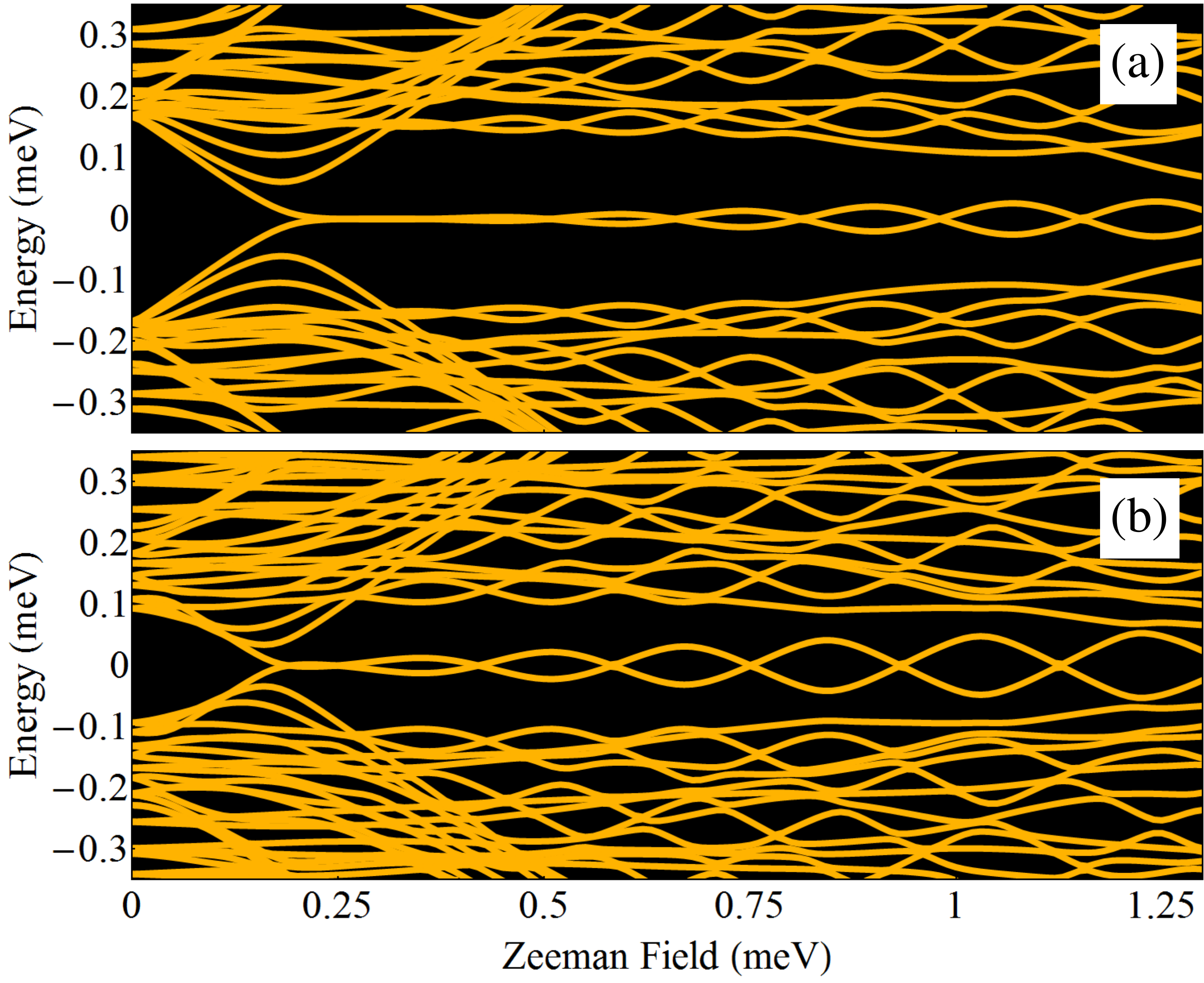}
\end{center}
\vspace{-4mm}
\caption{(Color online)  Low-energy spectrum as function of the applied Zeeman field for the 2D structure represented schematically in panel (b) of  Fig. \ref{FIG1S} (i.e. junction near the end of the Majorana wire) and two different values of the chemical potential (same as Fig. \ref{FIG2S}): (a) chemical potential near the bottom of the second confinement induced band ($\mu=3.6~$meV) and (b) chemical potential near the bottom of the third band ($\mu=8.1~$meV). Note the absence of any additional, junction-induced low-energy mode.}
\label{FIG3S}
\vspace{-5mm}
\end{figure}
%%%%%%%%%%%%%%%%%%%%%%	
%%%%%%%%%%%%%%%%%%%%%%%%%%%%%%%%

A possible solution is to place the junction near (or at) the end of the wire, as illustrated schematically in panel (b) of Fig. \ref{FIG1S}. The dependence of the corresponding low-energy spectrum on the Zeeman field is shown in Fig. \ref{FIG3S} (for the same two values of the chemical potential as in Fig. \ref{FIG2S}). Remarkably, there is no additional low-energy mode induced by the junction. We emphasize that in both scenarios illustrated in Fig. \ref{FIG1S} the junction can generate a strong perturbation. However, when the junction is near the middle of the wire (or more generally, far from the ends), this perturbation can induce trivial ABSs. By contrast, when the junction is placed at or near the end of the wire (within a distance smaller than the characteristic Majorana localization length scale $\xi$), it can only locally modify the Majorana wave function, but it does not affect its energy and, more importantly, it does not induce additional low-energy trivial ABSs. We note that this property holds under the assumption that the junction is ``sharp-enough'', i.e., it does not generate some smooth confinement  over length scales  larger than $\xi$). We note that a sharp junction can be seen as just a ``reconfiguration'' of the wire's end. By contrast, a smooth junction effectively extends the wire. 
Intuitively, when the junction is sharp, there is not enough ``room'' at the end of the wire for three low-energy Majorana modes (the MZM and the two Majoranas corresponding to a spurious ABS); two of the strongly overlapping modes acquire an energy of the order of the gap, while the MZM remains at zero energy.

We conclude that the optimal solution for engineering junctions between Majorana wires and trivial superconductors is to place them at or near the ends of the wire.
Finally, we note that our treatment of the junction between a Majorana wire and a trivial superconductor is equivalent to imposing a local boundary condition on the Majorana wire that is different from the boundary condition away from the junction. In fact, one of the main points of our study is that   junctions (of all kinds) modify locally the boundary conditions for the Majorana wire, which may cause perturbations strong enough to induce spurious low-energy states.

%%%%%%%%%%%%%%%%%%%%%%%%%%%%%%%%
%%%%%%%%%%%%%%%%%%%%%%%%%%%%
\begin{figure}[t]
\begin{center}
\includegraphics[width=0.48\textwidth]{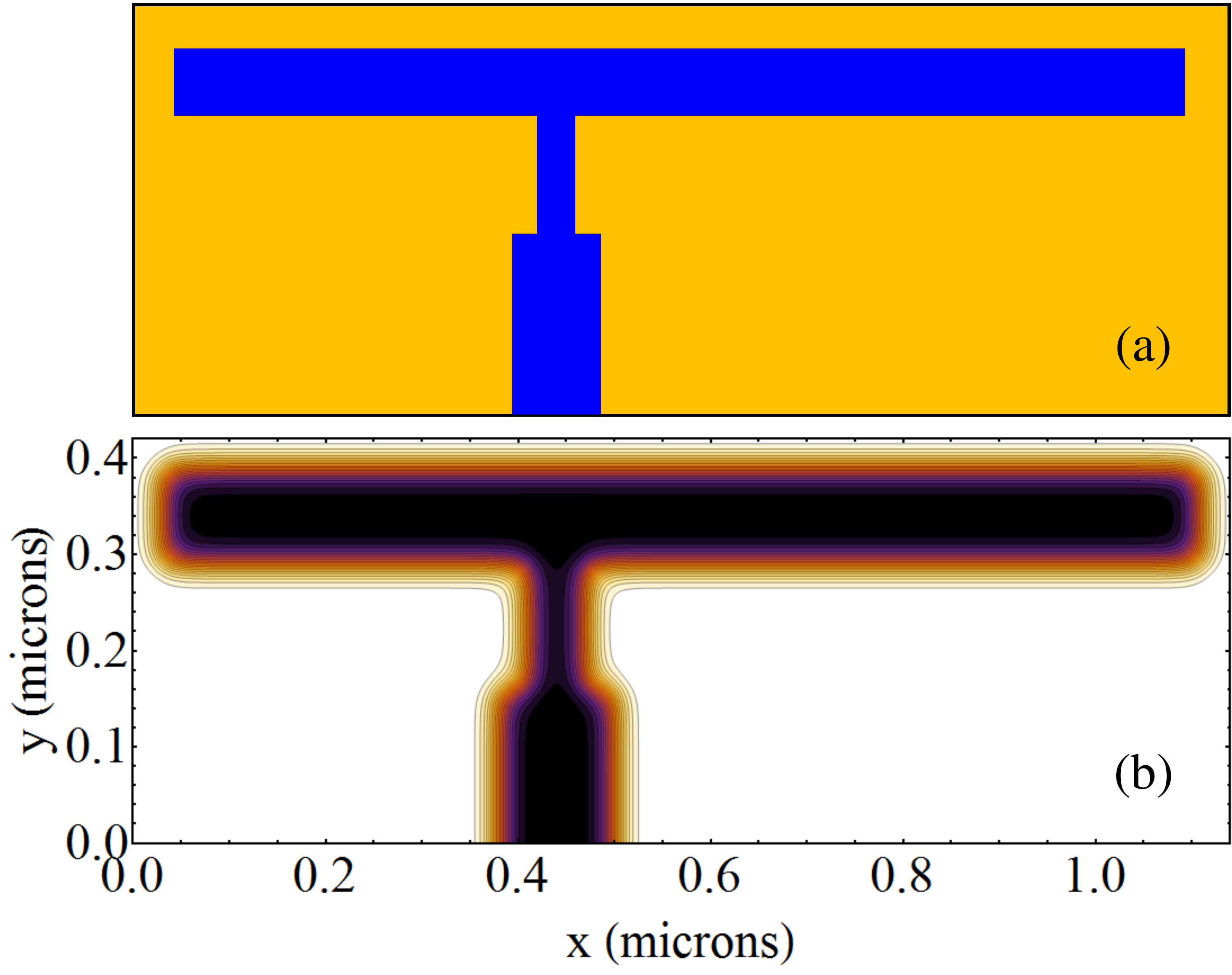}
\end{center}
\vspace{-2mm}
\caption{(Color online) (a) Schematic representation of a Majorana island grounded using a T-type wire junction. Note the constriction of the passive wire in the vicinity of the junction and the manifest asymmetry of the structure. (b) Effective confining potential corresponding to the device shown in panel (a).}
\label{FIG4S}
\vspace{-3mm}
\end{figure}
%%%%%%%%%%%%%%%%%%%%%%	
%%%%%%%%%%%%%%%%%%%%%%%%%%%%%%%%

Finally, we consider the most  straightforward solution for grounding a topological superconductor island in a 2D structure: using a  standard T-type wire junction. In general, this induces low-energy ABSs and, therefore, cannot be a valid solution for a topological quantum device. However, if we are interested in demonstrating correlated differential conductance features when tunneling into both ends of the wire, this may represent a possible approach. More specifically, one does not have to eliminate the junction-induced ABSs, but only to ensure that they have a low-enough overlap with the Majorana modes. We suggest a geometry similar to that shown in Fig. \ref{FIG4S}. There are two important features. First, the passive wire has a constriction at the junction with the Majorana wire.  This is to ensure that the junction-induced perturbation is as localized as possible.  Second, the junction is not in the middle of the Majorana wire. This is to ensure that the possible correlations of the differential conductance measured at the opposite ends reflect a property of the Majorana modes, rather than some ``artificial'' symmetry built into the device.  Finally, we note that the applied in-plane Zeeman field is perpendicular to the passive wire, hence parallel to the effective spin-orbit coupling field. Consequently, the passive wire becomes gapless, which is perfectly fine when using it as a contact.

\begin{figure}[t]
\begin{center}
\includegraphics[width=0.48\textwidth]{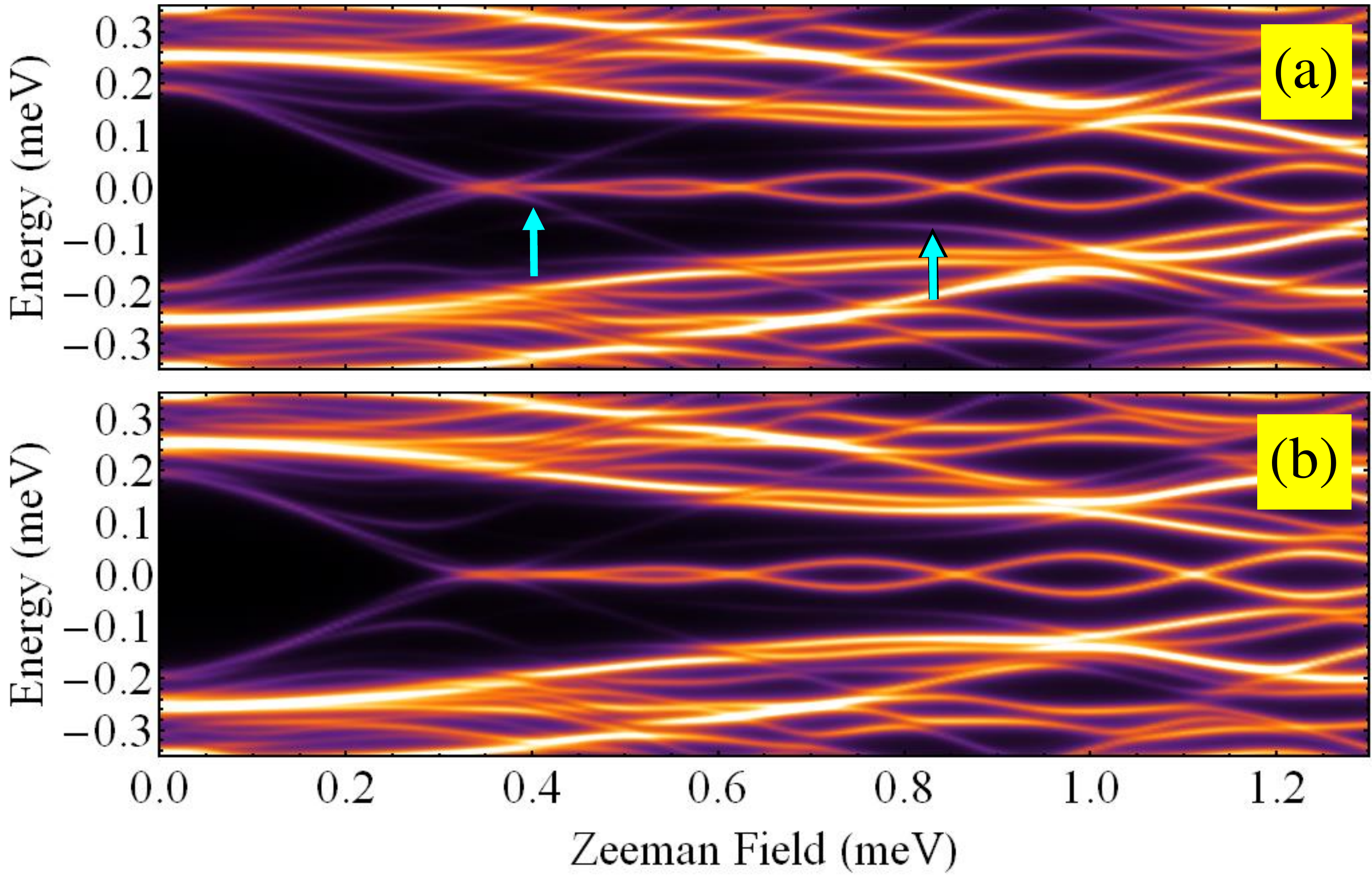}
\end{center}
\vspace{-4mm}
\caption{(Color online) Low-energy LDOS as a function of the Zeeman field for the 2D structure shown in Fig. \ref{FIG4S}. (a) LDOS at the left end of the Majorana wire. (b) LDOS at the right end of the Majorana wire. Note that the Majorana signatures at the two end of the wire are correlated. The arrows in panel (a) indicate (faint) signatures of the junction-induced ABS mode.}
\label{FIG5S}
\vspace{-3mm}
\end{figure}

The local density of states (LDOS) at the ends of the (grounded)  Majorana wire is shown in Fig. \ref{FIG5S}. Note that Majorana features revealed by the LDOS at the opposite ends are correlated. The perturbation generated by the junction induces a low-energy ABS, but its overlap with the Majorana modes is weak and does not affect the nonlocal correlations. Weak signatures of the ABS are marked by the arrows in panel (a). Note that a strong perturbation that effectively ``cuts'' the Majorana wire into two pieces is expected to destroy the correlations, unless the two segments are identical. This scenario makes clear the importance of constructing asymmetric junctions, rather than placing the junction in the middle of the Majorana wire. If the wire has a strong charged impurity or other type of intrinsic strong perturbation that divides it effectively into two segments, the MZM correlations are destroyed, since each segment has now two MZMs localized at its ends, completely suppressing any correlations between the two MZMs at the ends of the original wire. Such harmful situations are presumably avoided in the disorder-free ballistic wires being used for Majorana studies in the current experiments, but an explicit experimental verification is necessary. 

%%%%%%%%%%%%%%%%%%%%%%%%%%%%%%%%
%%%%%%%%%%%%%%%%%%%%%%%%%%%%
%%%%%%%%%%%%%%%%%%%%%%	
%%%%%%%%%%%%%%%%%%%%%%%%%%%%%%%%

\section{Concluding remarks}

We have studied the low-energy sub-gap electronic structure of Majorana wires in the presence of junctions with proximitized semiconductor nanowires and trivial superconductors. We found that the presence of a junction generally induces spurious low-energy sub-gap states that can destroy the topological protection of the Majorana subspace or corrupt the low-energy spectral features, when the additional wire is used as a tunneling probe. Uncontrolled spurious zero-energy (or even low-enough-energy) sub-gap features are a potential  source of errors for topological quantum computation, including measurement-only protocols. The source of the perturbation responsible for the spurious states is the nonuniform transverse confinement that characterizes the system in the presence of the junction. We show that a possible solution to this problem is to engineer junctions in 2D electron systems hosted by semiconductor-superconductor heterostructures. We also propose a specific 2D device that enables multiprobe tunneling experiments capable of providing position-dependent spectroscopy. In addition, we find that the optimal solution for building junctions with trivial superconductors is to place them near the end of the wire and we show that  the poor man's solution for grounding a topological superconductor island using a standard T-junction could work in certain conditions. Although the junction induces additional low-energy states, these states may be ``invisible''  in an experiment involving tunneling from both ends, which will show correlated features. However, if correlations are not observed, this may be either an intrinsic property of the Majorana wire, or a result of the perturbation induced by the junction.

A few final remarks are warranted. (1) The perturbation induced by a T-junction can be minimized if the Majorana wire is only partially covered by the superconductor and the potential created by a gate running parallel to the wire [see Fig. \ref{FIG1}(a)] pushes the electrons away from the side containing the junction. The resulting ``soft'' effective confinement is similar to that shown in Fig. \ref{FIG2}(d). 
(2) Using a structure like that shown in Fig. \ref{FIG4S} for a two-end conductance measurement would be particularly relevant for relatively short wires, where at least one Majorana energy splitting (if not splitting oscillations) can be observed. In a long wire with a featureless Majorana signature the only possible correlation involves the onset field associated with the emergence of the ZBP. However, observing such a correlation does not eliminate the existence of other MZMs being localized throughout the wire (but far enough from the ends). In fact, this is a generic problem involving the two-end measurement (regardless of what technical solution is used for grounding the parent SC). (3) The best way to conclusively demonstrate that the only (nearly) zero-energy modes are those localized near the ends of the wire is to have access to position-dependent spectral information. As shown in this study, this can be obtained using a 2D structure like that shown in panel (b) of Fig. 1. The Majorana wire can be grounded using a T-wire junction, while the probing ``fingers'' are placed on the other side of the wire. If a junction-induced ABS emerges, it will be detected by the nearby probe(s). However, if the Majorana wire has a quality that is consistent with the requirements of topological quantum devices, there should be no low-energy modes other than the MZMs localized at the ends of the wire and (possibly) the ABS induced by the junction. Finally, if the Majorana wire is grounded using a junction with a trivial superconductor placed near one of the ends [e.g., as shown in Fig. \ref{FIG1S} (b)] and the probing ``fingers'' are placed along the wire, demonstrating topologically-protected  MZMs implies demonstrating that  the only low-energy states emerge above a critical field associated with a minimum of the bulk gap and are localized at the ends of the wire.

\vspace{1mm}

\noindent {\em Note added}: A preprint posted recently, Vaitiekenas {\em et al}., arXiv:1710.04300, reports a gap-reopening feature that is consistent with the presence of Andreev bound states localized near the end of the wire.

\vspace{-3mm}

\begin{acknowledgments}
This  work  is  supported  by  Microsoft,  Laboratory for Physical Sciences, NSF DMR-1414683, and WV HEPC/dsr.12.29. TDS acknowledges helpful conversations with Chetan Nayak and Fabrizio Nichele.
\end{acknowledgments}

\bibliography{REFERENCES}

\end{document}